\documentclass[aps,pra,amsmath,amssymb,letterpaper,twocolumn]{revtex4}

\usepackage{hyperref}
\usepackage{graphicx}
\usepackage{color}

\begin{document}

\title{Approaching the Full Configuration Interaction Low-Energy
  Spectrum from an Arbitrary Reference Subspace}

\author{Carlos A. Jim\'enez-Hoyos}
\email{cjimenezhoyo@wesleyan.edu}
\affiliation{Department of Chemistry, Wesleyan University, Middletown,
  CT, 06459}

\date{\today}

\begin{abstract}
In a previous work (\href{arxiv.org/2010.02027}{arXiv:2010.02027}) we
showed how the full configuration interaction (FCI) ground state
energy can be obtained as a functional of an arbitrary reference
wavefunction by means of a gradient descent or quasi-Newton
algorithm. Here, we extend this approach and consider the optimization
of the low-energy subspace of the Hamiltonian from an arbitrary
reference subspace. The energies along the optimization path are
obtained in terms of transition matrix elements among the states in
the reference subspace. We show an application of the algorithm with a
reference subspace constructed from a non-orthogonal configuration
interaction (NOCI) formalism to describe the avoided crossing in LiF
and the low-lying singlet and triplet spectrum of formaldehyde.
\end{abstract}

\maketitle

\section{Introduction}

Exact solutions to the electronic Schr\"odinger equation can only be
obtained, in closed form, for very small chemical systems. Therefore,
most quantum chemical calculations aim to reproduce, as closely as
possible, full configuration interaction (FCI) solutions
\cite{szabo_ostlund}, where the electronic Scr\"odinger equation is
projected onto a basis of $N$-electron wavefunctions constructed from
antisymmetrized products of some suitable one-particle basis. In
trying to reproduce FCI solutions, one main goal is to reduce the
computational effort as much as possible. A number of methods that can
yield arbitrarily accurate approximations to the ground state
wavefunction are known, but the number of methods available to target
excited electronic states is more limited.

In a previous work \cite{jimenez2020}, hereafter referred to as paper
I, we discussed gradient descent and quasi-Newton algorithms to reach
the FCI ground state wavefunction starting from an arbitrary reference
state $|0 \rangle$. The central goal of that paper was to define
systematic approximations to the ground state wavefunction,
characterized by the number of steps taken in the algorithm. Along
with that goal, a key message of that work was that the energies along
the optimization path can be written in terms of matrix elements of
$|0 \rangle$ and, therefore, an explicit vector representation of the
wavefunction need not be built or stored.  The present work is a
generalization of paper I where we now consider the optimization of a
low-energy subspace starting from a reference set of wavefunctions.

Some of the advantages of avoiding an explicit vector representation
of the wavefunction are:
\begin{itemize}
  \item Given some number $k$ of steps to take along the optimization
    path, the required matrix elements can be evaluated in polynomial
    time for a number of common wavefunctions.
  \item There is no need to store wavefunctions as vectors that have a
    dimension of the Hilbert space. This is turn allows calculations
    in systems with much larger Hilbert space dimensions than would
    otherwise be possible.
  \item It is possible to consider reference states for which a
    simple, {\em a priori}, construction of the orthogonal complement
    is not readily available.
\end{itemize}
All of those advantages are still relevant to the present work where
we work with a reference subspace of wavefunctions rather than a
single state $|0 \rangle$.

In this work, we focus our application of the algorithm to
non-orthogonal configuration interaction (NOCI) expansions
\cite{sundstrom2014,thom2018} for the low-energy spectrum of molecular
systems. (We stress, however, that the algorithm is applicable to
other types of wavefunctions.) Here, we use NOCI in a broad sense to
refer to ground or/excited states written as linear combinations of
generally non-orthogonal ($\langle \Phi_q | \Phi_p \rangle \neq 0$)
determinants:
\begin{equation}
  |k \rangle = \sum_q f^k_q |\Phi_q \rangle \nonumber,
\end{equation}
where $f$ are some linear coefficients determined from the
corresponding generalized eigenvalue problem.  Moreover, we also
consider similar expansions written in terms of symmetry-projected
Slater determinants \cite{jimenez2012} of the form
\begin{equation}
  |k \rangle = \sum_q f^k_q \hat{P} \, |\Phi_q \rangle \nonumber.
\end{equation}
where $\hat{P}$ is a projection operator that restores some symmetry
of the Hamiltonian. Some of our recent work \cite{jimenez2019} has
shown that NOCI expansions based on symmetry-projected configurations
can yield a qualitatively correct low-energy spectrum of molecular
systems. Therefore, the goal of the present paper is to explore how
that reference subspace can be evolved, using gradient descent and
quasi-Newton algorithms, to yield the exact FCI low-energy spectrum.

The rest of this manuscript is organized as follows. In
Sec. \ref{sec:theory} we describe the optimization target as well as
the parametrization we use in carrying out the optimization. We then
provide details of gradient descent (\ref{sec:gradient}) and
quasi-Newton (\ref{sec:qnewton}) optimization algorithms, providing
explicit expressions for the first few iterations. In
Sec. \ref{sec:results} we discuss the application of the method in a
H$_4$ ring, in the avoided crossing of LiF, and in the low-energy
spectrum of formaldehyde. Finally, in Sec. \ref{sec:conclusions} we
provide some closing remarks.

\section{Theory}
\label{sec:theory}

We consider the optimization of the lowest-energy $q$ FCI
states. Our optimization target is the state-averaged energy
\begin{equation}
  E_\mathrm{SA} = \frac{1}{q} \left( \varepsilon_0 + \varepsilon_1 +
  \ldots + \varepsilon_{q-1} \right),
\end{equation}
where $\varepsilon_0$ and $\varepsilon_{q-1}$ correspond to the
Hamiltonian ground and $(q-1)$-th excited eigenvalues, respectively.
Naturally, a minimum of $E_\mathrm{SA}$ coincides with convergence of
the entire low-energy subspace of dimension $q$.

Without loss of generality, we assume that a set of orthonormal
reference wavefunctions is available: $\{|0\rangle, |1\rangle, \ldots,
|q-1\rangle\}$. We use an exponential, non-Hermitian parametrization
to build states of the form
\begin{align}
  |\Psi^k \rangle &= \, \exp (\hat{Z}^k) |k \rangle, \label{eq:Z}
  \\ \hat{Z}^k &= \, \sum_x Z_x^k |x \rangle \langle k|,
\end{align}
where $|x \rangle$ labels an orthonormal state in the orthogonal
complement of the reference subspace. The function to be optimized is
then $E_\mathrm{SA} [Z]$, with $Z$ being the coefficients of the
$\hat{Z}^k$ operators in Eq. \ref{eq:Z}. When convergence is reached,
the FCI low-energy spectrum can be recovered from the solution to the
generalized eigenvalue problem $\mathcal{H} \, C = \mathcal{S} \, C \,
\varepsilon$, with
\begin{align}
  \mathcal{S}_{kl}(Z) &\equiv \, \langle k| \exp(\hat{Z}^{k,\dagger})
  \exp(\hat{Z}^l) |l \rangle, \\
  \mathcal{H}_{kl}(Z) &\equiv \, \langle k| \exp(\hat{Z}^{k,\dagger}) H
  \exp(\hat{Z}^l) |l \rangle.
\end{align}
Nonetheless, the low-energy spectrum can be determined at any point in
the optimization from the solution of the corresponding generalized
eigenvalue problem. The matrix elements of $\mathcal{S}$ and
$\mathcal{H}$, as a function of $Z$, are given by
\begin{align}
  \mathcal{S}_{kl}(Z) &= \, \delta_k^l + Z^x_k Z^l_x, \\
  \mathcal{H}_{kl}(Z) &= \, H_k^l + H_k^x Z^l_x + Z^x_k H_x^l +
   Z^x_k H_x^y Z^l_y,
\end{align}
where $H_\alpha^\beta = \langle \alpha | H | \beta \rangle$, Einstein
summation is implied and the indices $x,y$ run only over the
orthogonal complement of the reference subspace. We have assumed that
the reference wavefunctions $\{ |k \rangle \}$ are real and therefore
we use real coefficients $Z$, as we do throughout this work. While the
state-averaged energy can be expressed in terms of the Hamiltonian
eigenvalues, it can also be written as
\begin{equation}
  E_\mathrm{SA}[Z] = \frac{1}{q} \mathrm{Tr} \Big( \mathcal{H}(Z) \,
  \mathcal{S}^{-1}(Z) \Big).
\end{equation}
The gradient of $E_\mathrm{SA}$ with respect to $Z$, evaluated at
$Z=Y$ is given by
\begin{align}
  g^k_x &\equiv \, \left. \frac{\partial E_\mathrm{SA} [Z]}{\partial
    Z^x_k} \right|_{Z=Y} \nonumber \\
  &= \, \frac{2}{q} \sum_{l} \left( H_x^l +
  H_x^y \, Y^l_y \right) \Big( \mathcal{S}^{-1} (Y) \Big)_{lk} \nonumber \\
  &-\,  \frac{2}{q} \sum_l
  Y^l_x \Big( \mathcal{S}^{-1} (Y) \, \mathcal{H} (Y) \, \mathcal{S}^{-1} (Y)
  \Big)_{lk}.
\end{align}

For convenience, we shall introduce the matrices
\begin{align}
  (F_1)_k^l &= \, H_k^l, \\
  (F_2)_k^l &= \, H_k^x H_x^l, \\
  (F_3)_k^l &= \, H_k^x H_x^y H_y^l.
\end{align}
Note that all elements in $F_1$, $F_2$, \ldots can be evaluated in
terms of matrix elements (or transition matrix elements) from the
reference subspace. For instance,
\begin{align}
  (F_2)_k^l &= \, \langle k| H^2 |l \rangle -
  \sum_m \langle k| H |m \rangle (F_1)_m^l, \\
  (F_3)_k^l &= \, \langle k| H^3 |l \rangle -
  \sum_m \langle k| H^2 |m \rangle (F_1)_m^l \nonumber \\
  &- \, \sum_m \langle k| H |m \rangle (F_2)_m^l.
\end{align}

\subsection{Gradient Descent}
\label{sec:gradient}

We begin at $Z_0 = 0$ with $|\Psi_0^k \rangle = |k\rangle$. Naturally,
$E_\mathrm{SA}^0 = 1/q \sum_k H_k^k$. The gradient at $Z_0$ is
\begin{align}
  (g_0)^k_x &= \, H_x^l (\alpha_0)^k_l,
\end{align}
with $(\alpha_0)^k_l = 2/q \, \delta^k_l$.

Just as in paper I, we shall consider a full line search along
$-g_0$. The state-averaged energy, as a function of the step size
$\sigma$, is given by
\begin{equation}
  E_\mathrm{SA}^1[-\sigma g_0] = \frac{1}{q} \mathrm{Tr} \Big(
  \mathcal{H}_1(\sigma) \, \mathcal{S}_1^{-1} (\sigma) \Big),
\end{equation}
with
\begin{align}
  \mathcal{H}_1 (\sigma) &= \, F_1 - \sigma \alpha_0 \, F_2 - \sigma
  F_2 \alpha_0 + \sigma^2 \alpha_0 \, F_3 \, \alpha_0, \\
  \mathcal{S}_1 (\sigma) &= \, 1 + \sigma^2 \alpha_0 \, F_2 \,
  \alpha_0.
\end{align}
Note that $\mathcal{H}_1$ and $\mathcal{S}_1$ can be assembled from
matrix elements in the reference subspace, as decribed
above. Therefore, $E_\mathrm{SA}^1$ is itself a functional of the
reference subspace that can be determined after evaluation of $F_1$,
$F_2$, and $F_3$.  A closed-form solution for $\sigma_\ast$ that
minimizes $E_\mathrm{SA}^1[-\sigma g_0]$ is possible, but it is easier
in practice to carry out the minimization numerically. With
$\sigma_\ast$ available, the states $|\Psi_1^k \rangle$ can be written
as
\begin{equation}
  |\Psi_1^k \rangle = e^{-\sigma_\ast (g_0)^k} |k \rangle.
\end{equation}

We can now attempt a second step. The gradient at $Z_1 = -\sigma_\ast
g_0$ is
\begin{equation}
  (g_1)^k_x = H_x^l (\alpha_1)^k_l + H_x^y H_y^l (\beta_1)^k_l ,
  \label{eq:g1}
\end{equation}
with
\begin{align}
  (\alpha_1)^k_l &= \, \frac{2}{q} \Big( \mathcal{S}^{-1}_{1\ast}
  + \sigma_\ast \alpha_0 \, \mathcal{S}^{-1}_{1\ast}
  \mathcal{H}_{1\ast} \mathcal{S}^{-1}_{1\ast}
  \Big)_{lk}, \\
  (\beta_1)^k_l &= \, \frac{-2}{q} \Big( \sigma_\ast \alpha_0 \,
  \mathcal{S}^{-1}_{1\ast} \Big)_{lk},
\end{align}
where $S_{1\ast} \equiv S_1(\sigma_\ast)$ and $H_{1\ast} \equiv
H_1(\sigma_\ast)$.

Considering a line search along $-g_1$ with step size $\tau$, we can
write the state-averaged energy as a function of $\tau$
\begin{equation}
  E_\mathrm{SA}^2[-\sigma_\ast g_0 - \tau g_1] = \frac{1}{q}
  \mathrm{Tr} \Big( \mathcal{H}_2(\tau) \, \mathcal{S}_2^{-1}
  (\tau) \Big),
  \label{eq:e2}
\end{equation}
with
\begin{align}
  \mathcal{H}_2 (\tau) = \mathcal{H}_{1\ast} &-\, \tau \alpha_1 F_2 -
  \tau F_2 \alpha_1 - \tau \beta_1 F_3 - \tau F_3 \beta_1 \nonumber \\
  &+\, \tau^2 \alpha_1 F_3 \alpha_1 + \tau^2 \beta_1 F_5 \beta_1 \nonumber \\
  &+\, \tau^2 \alpha_1 F_4 \beta_1 + \tau^2 \beta_1 F_4 \alpha_1 \nonumber \\
  &+\, \tau \sigma_\ast \alpha_0 F_3 \alpha_1 + \tau \sigma_\ast
  \alpha_1 F_3 \alpha_0 \nonumber \\
  &+\, \tau \sigma_\ast \alpha_0 F_4 \beta_1 + \tau
  \sigma_\ast \beta_1 F_4 \alpha_0, \\
  \mathcal{S}_2 (\tau) = \mathcal{S}_{1\ast}
  &+\, \tau^2 \alpha_1 F_2 \alpha_1 + \tau^2 \beta_1 F_4 \beta_1 \nonumber \\
  &+\, \tau^2 \alpha_1 F_3 \beta_1 + \tau^2 \beta_1 F_3 \alpha_1 \nonumber \\
  &+\, \tau \sigma_\ast \alpha_0 F_2 \alpha_1 + \tau \sigma_\ast
  \alpha_1 F_2 \alpha_0 \nonumber \\
  &+\, \tau \sigma_\ast \alpha_0 F_3 \beta_1 + \tau \sigma_\ast
  \beta_1 F_3 \alpha_0.
\end{align}
Note that $\mathcal{H}_2$ and $\mathcal{S}_2$ can be assembled with
$F_1$, $F_2$, \ldots, $F_5$ available. Therefore, $E_\mathrm{SA}^2$ is
still a functional of the reference subspace. Let $\tau_\ast$ be the
minimizer of $E_\mathrm{SA}^2[-\sigma_\ast g_0 - \tau g_1]$; the
states $|\Psi_2^k \rangle$ can then be written as
\begin{equation}
  |\Psi_2^k \rangle = e^{-\sigma_\ast (g_0)^k - \tau_\ast (g_1)^k} |k
  \rangle.
\end{equation}

If a third step is attempted, by simple inspection one can readily
realize that the gradient at $Z_2 = -\sigma_\ast g_0 - \tau_\ast g_1$
takes the form
\begin{align}
  (g_2)^k_x &= \, H_x^l (\alpha_2)^k_l + H_x^y H_y^l (\beta_2)^k_l +
  H_x^y H_y^z H_z^l (\gamma_2)^k_l.
\end{align}
The corresponding $E_\mathrm{SA}^3$ would also be a functional of the
reference subspace that can be assembled from $F_1$, $F_2$, \ldots,
$F_7$. Subsequent steps require the evaluation of higher order $F_k$
matrices.

\subsection{Quasi-Newton}
\label{sec:qnewton}

We now consider a quasi-Newton approach \cite{nocedal} in order to
improve the rate of convergence of the state-averaged energy. At each
step along the optimization, the search direction $p_k$ is determined
from $B_k p_k = -g_k$, rather than setting $p_k = -g_k$ as in gradient
descent. We perform a full line search along $p_k$ as in gradient
descent. Following our previous work, we choose to set $B_0 = I$ as
this allows us to fully define the quasi-Newton method as a functional
of the reference subspace.

With the choice $B_0 = I$, the first step coincides with that from
gradient descent and $E^1_\mathrm{SA}$ remains unchanged. While the
gradient $g_1$ is the same as in gradient descent (see
Eq. \ref{eq:g1}), the search direction $p_1$ is determined from $p_1 =
-B_1 g_1$, with $B_1$ constructed using a quasi-Newton update formula.

As shown in appendix \ref{sec:bfgs}, $p_1$ determined from a
Broyden-Fletcher-Goldfarb-Shanno (BFGS)
\cite{broyden,fletcher,goldfarb,shanno} update formula, takes the form
\begin{equation}
  (p_1)^k_x = H_x^l (\alpha'_1)^k_l + H_x^y H_y^l (\beta'_1)^k_l ,
\end{equation}
with $\alpha'_1$ and $\beta'_1$ being some matrices that are
numerically different from $\alpha_1$ and $\beta_1$. Given that $p_1$
takes the same functional form as $g_1$, we conclude that
$E^2_\mathrm{SA}$ determined from the BFGS approach is also a
functional of $F_1$, \ldots, $F_5$. Namely, $E_2$ would take the same
form as Eq. \ref{eq:e2}, with $\alpha_1 \to \alpha'_1$ and $\beta_1
\to \beta'_1$ in the definitions of $\mathcal{H}_2(\tau)$ and
$\mathcal{S}_2(\tau)$. Further quasi-Newton steps can also be cast as
functionals of the reference subspace, with higher order $F_k$
matrices required.

\section{Results and Discussion}
\label{sec:results}

We proceed to discuss the application of the optimization algorithms
described above in a H$_4$ ring, in the avoided crossing of LiF, and
in the low-lying singlet and triplet spectrum of formaldehyde.

\subsection{H$_4$}

\begin{figure}[!htb]
  \includegraphics[width=8.5cm]{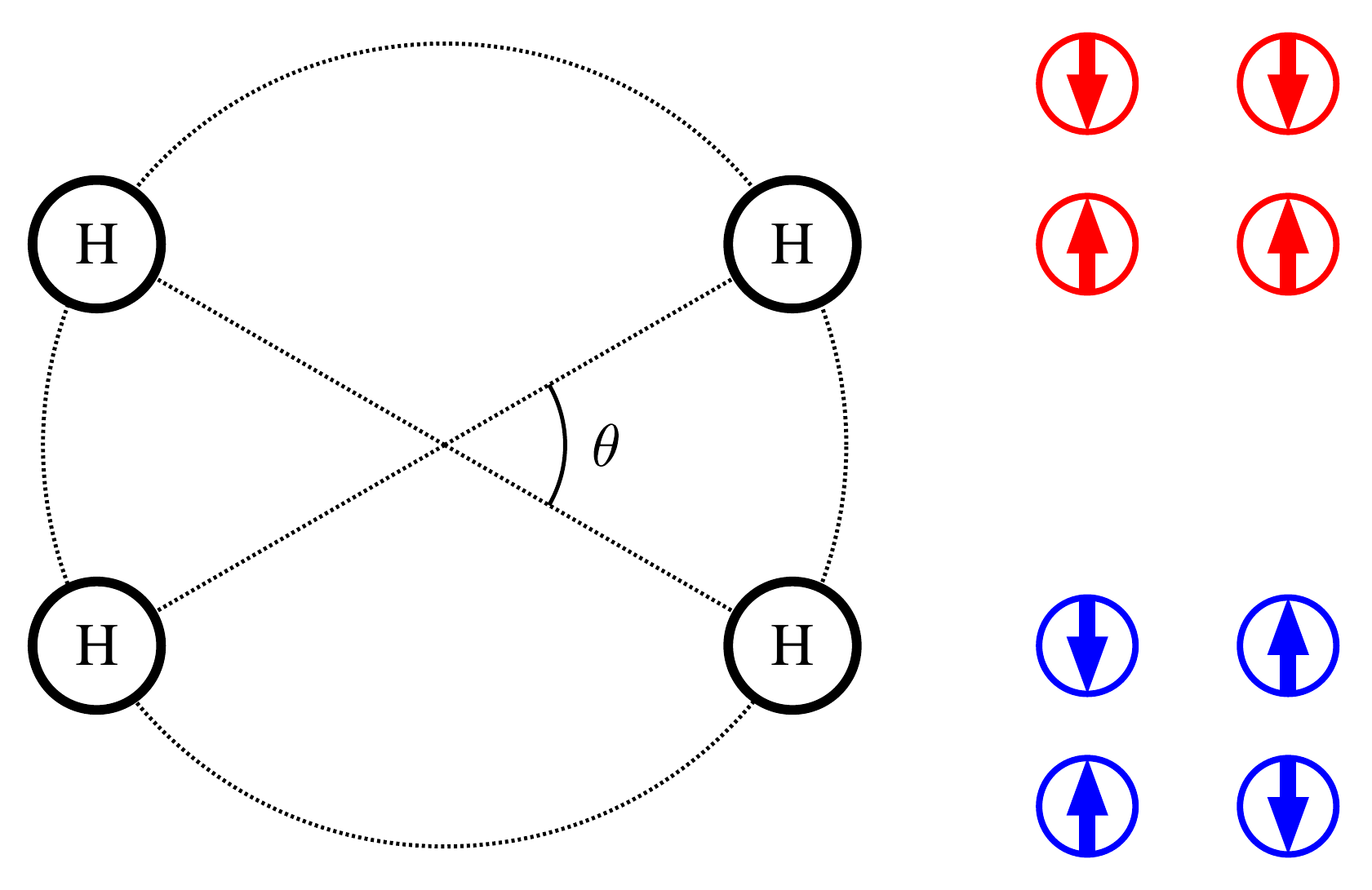}
  \caption{The H$_4$ system considered in paper I and introduced in
    Ref. \onlinecite{scuseria2017} consists of four H atoms placed
    along a ring of radius $r=3.3$~bohr, controlled by an angle
    $\theta$. To the right we show, with different colors, the spin
    arrangement in the two different UHF solutions
    considered. \label{fig:h4s}}
\end{figure}

We begin by revisiting the H$_4$ system discussed in paper I and shown
in Fig. \ref{fig:h4s}. While in paper I we focused on the ground
state, there are in fact two low-lying singlet states at large
$\theta$. Those two states can be reasonably well described using two
different unrestricted Hartree--Fock (UHF) configurations, whose
character is depicted in Fig. \ref{fig:h4s}.

\begin{figure*}[!htb]
  \includegraphics[width=8cm]{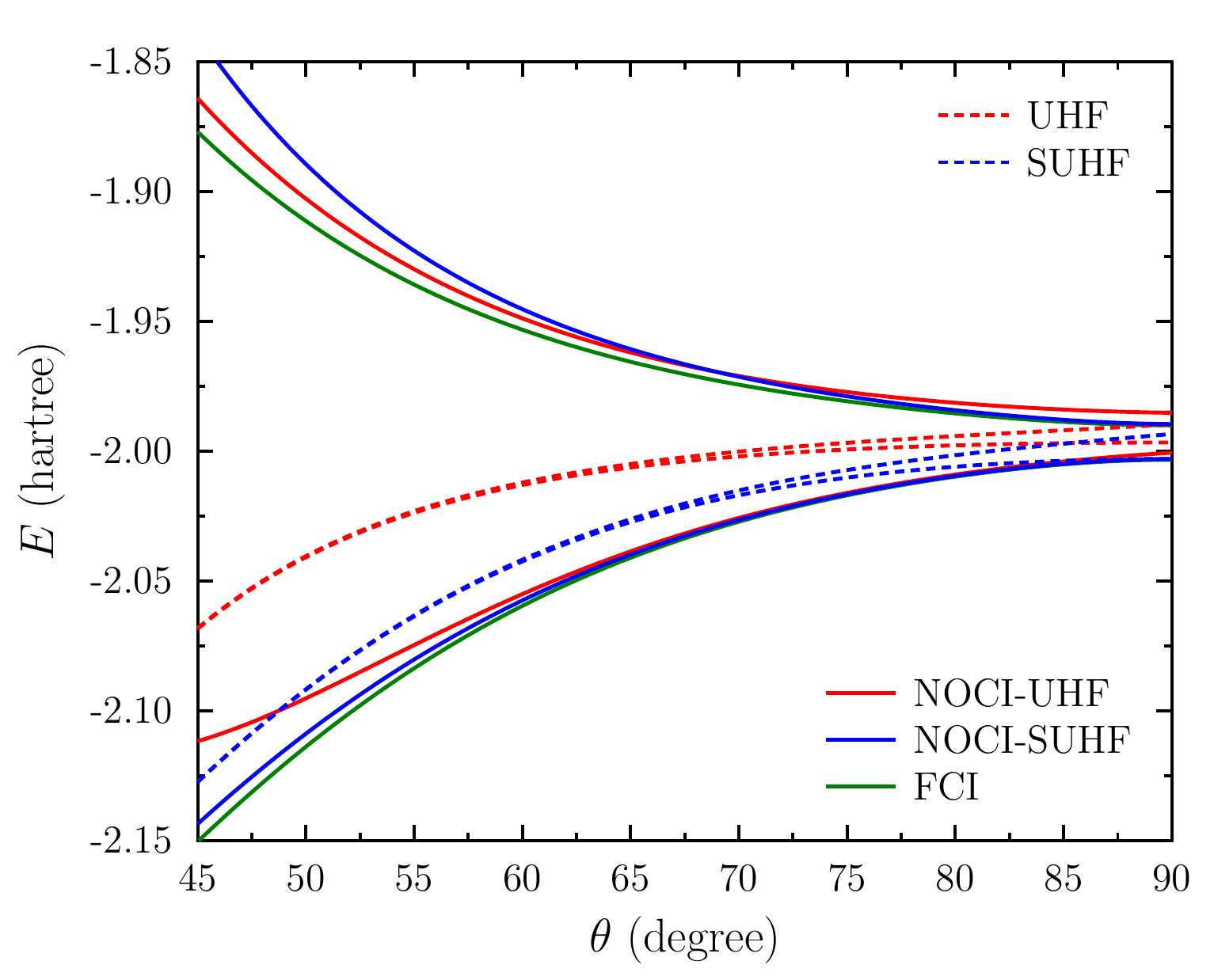}
  \hspace{0.2cm}
  \includegraphics[width=8cm]{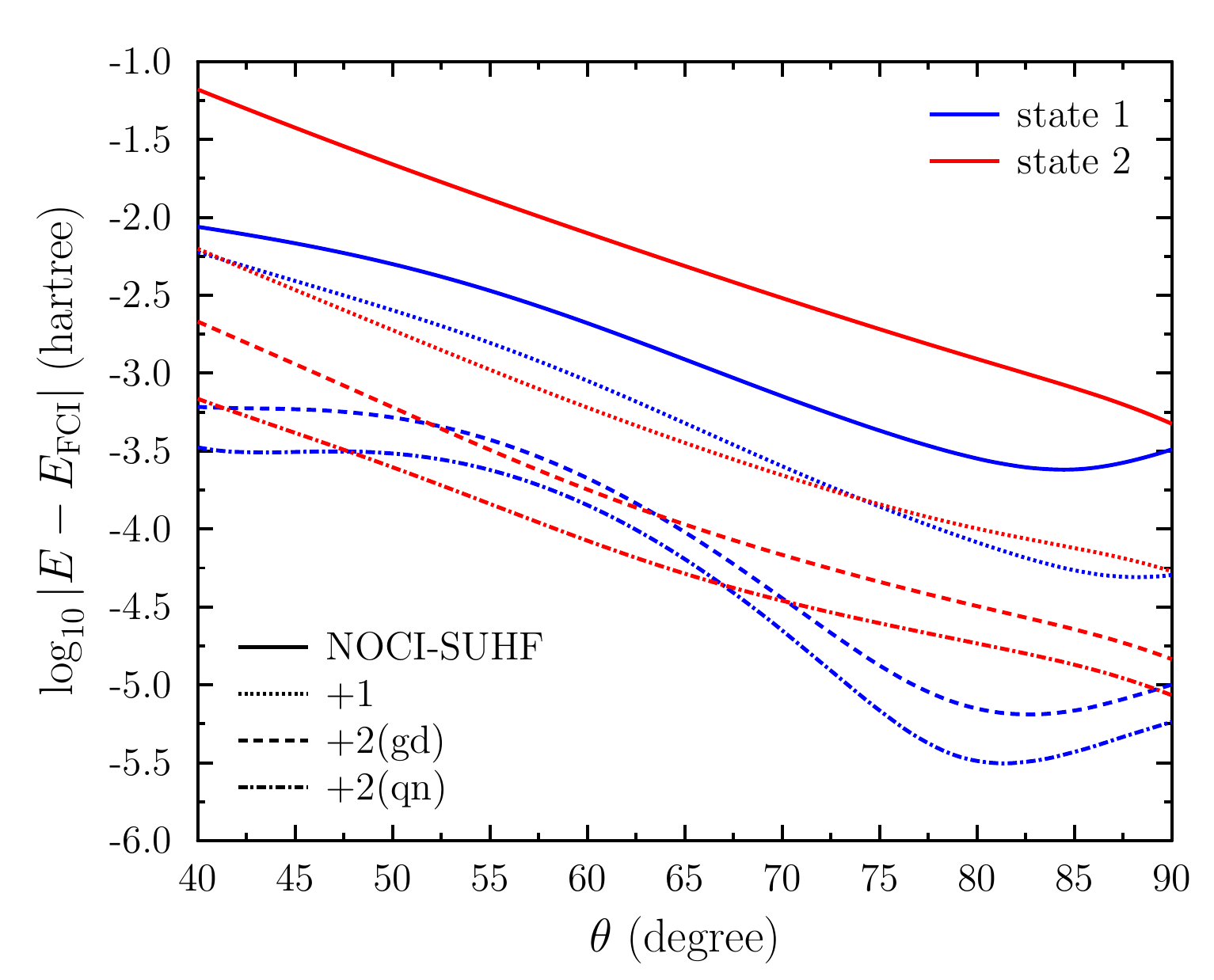}
  \caption{(Left) Energy (in hartree) of the two low-lying singlet
    states in the H$_4$ system of Fig. \ref{fig:h4s}, as a function of
    $\theta$, computed with NOCI-UHF, NOCI-SUHF, and FCI using the
    6-31G basis set. In addition, the energy of the two UHF solutions
    used in NOCI-UHF and the two SUHF solutions used in NOCI-SUHF is
    also displayed. Note that the two UHF solutions (and the two SUHF
    solutions) become nearly identical in energy for $\theta <
    70$~deg. (Right) Error in the energy, with respect to FCI, for the
    two low-lying singlet states. Results are shown with NOCI-SUHF and
    after 1 and 2 gradient descent (gd) or quasi-Newton (qn)
    steps. \label{fig:h4}}
\end{figure*}

We show in the left panel of Fig. \ref{fig:h4} the energy of the two
different UHF solutions, as well as the energy of the two
spin-projected UHF (SUHF) solutions that have a similar character. We
also show the energy obtained in NOCI-UHF where we use the two UHF
solutions plus their spin-flipped counterparts and consider the
resulting two eigenvectors with ``singlet'' character.\footnote{Those
  are the eigenvectors which are symmetric under flipping all of the
  spins.} We also show NOCI-SUHF curves obtained using the two
different SUHF solutions, as well as the two FCI lowest-lying singlet
states.

The right panel of Fig. \ref{fig:h4} shows the improvement to both
low-lying singlet states, as a function of $\theta$, after one or two
gradient descent (gd) or quasi-Newton (qn) steps have been taken,
starting from the NOCI-SUHF reference subspace. The energy of both
states is improved substantially even after just one iteration. Using
a qn algorithm yields better results after two steps than using a gd
algorithm, but in both cases the energy of both low-lying singlet
states is within a mHartree of the exact FCI results.

\begin{figure}[!htb]
  \includegraphics[width=8.5cm]{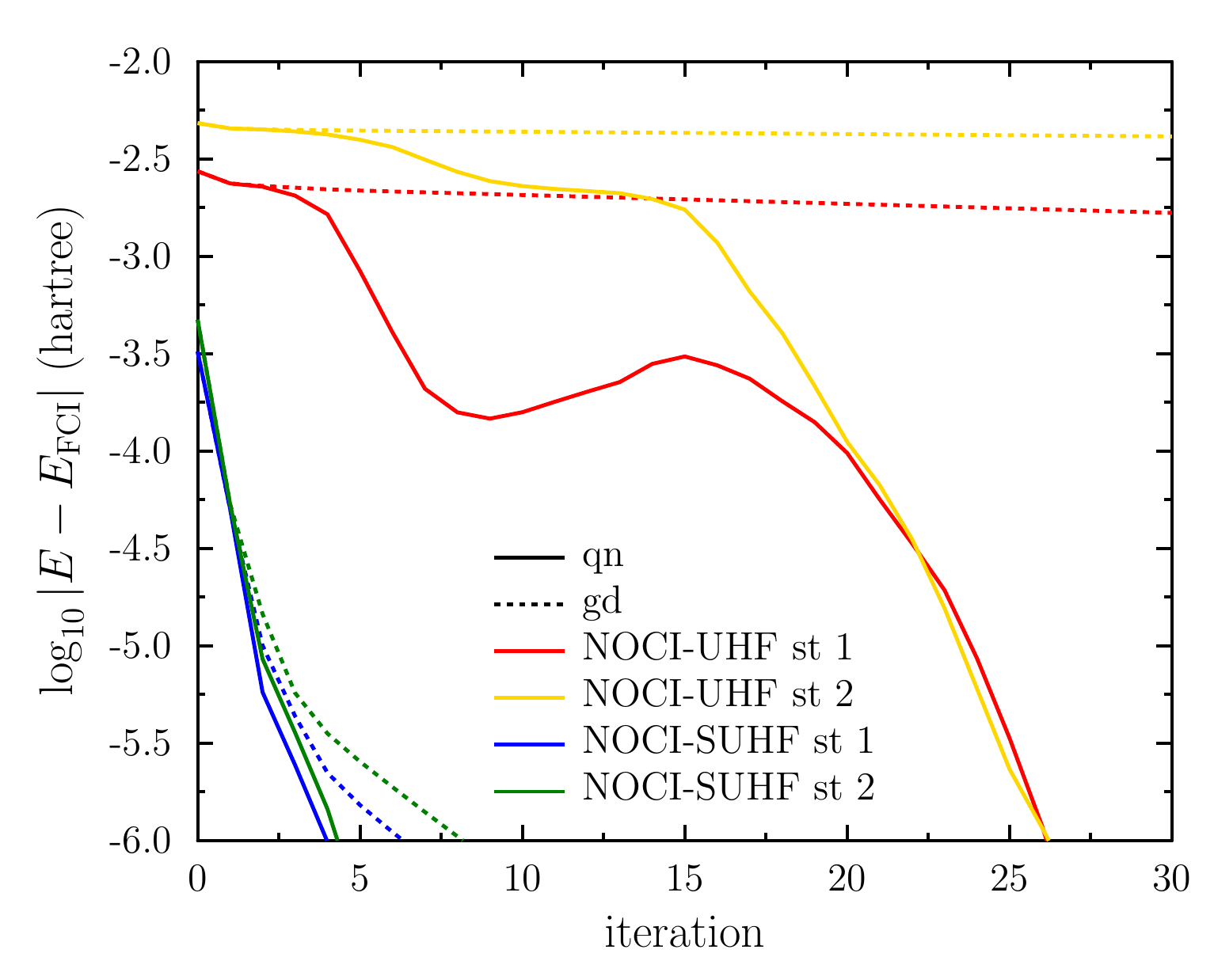}
  \caption{Convergence of the energy of the two low-lying singlet
    states as a function of iteration, at $\theta = 90$~deg using
    NOCI-UHF and NOCI-SUHF and gradient descent (gd) and quasi-Newton
    (qn) algorithms.
    \label{fig:conv}}
\end{figure}

We show in Fig. \ref{fig:conv} the convergence profile of the gd and
qn algorithms using the NOCI-UHF and NOCI-SUHF reference subspaces at
$\theta = 90$~deg. It is again evident that the qn algorithm reaches
convergence significantly faster than the gd algorithm, as
expected. The NOCI-SUHF reference subspace is significantly better
than the NOCI-UHF one and convergence (with $\mu$Hartree accuracy) is
reached after a handful of iterations. Using a qn algorithm, both in
the case of NOCI-UHF and NOCI-SUHF, convergence of the energy for both
low-lying states can be reached with a similar number of iterations.

\subsection{LiF Avoided Crossing}

We now consider the avoided crossing in the LiF potential energy
curve. A correct description of the avoided crossing requires a
multi-reference description \cite{bauschlicher1988}. In this case, we
work with a NOCI description, following our previous work
\cite{jimenez2019b}. Explicitly, we describe the two $^1 \Sigma$
states by using the restricted Hartree--Fock (RHF) determinant and the
symmetric linear combination of the UHF determinant and its
spin-flipped counterpart (technically, the later is not a spin singlet
state, but the most significant triplet contaminant has been
removed). The resulting NOCI states provide a qualitatively correct
description of the avoided crossing, but the quantitative aspects are
not quite correct. We have evaluated the LiF potential energy curve
using the same basis set as that in
Ref. \onlinecite{bauschlicher1988}, where frozen-core FCI results were
published.

We have calculated the correction in the state-averaged energy after 1
step starting from the NOCI reference subspace: sa-NOCI+1. Here, we
emphasize that this computation was done by evaluation of the $F_2$
and $F_3$ matrices, without an explicit vector representation of the
two states: in this case, the dimension of the FCI vector, in a basis
of $m_s = 0$ Slater determinants is $1.42 \times 10^{11}$, which would
render storage of the FCI vector impossible in most common
computational facilities.

\begin{figure*}[!htb]
  \includegraphics[width=8cm]{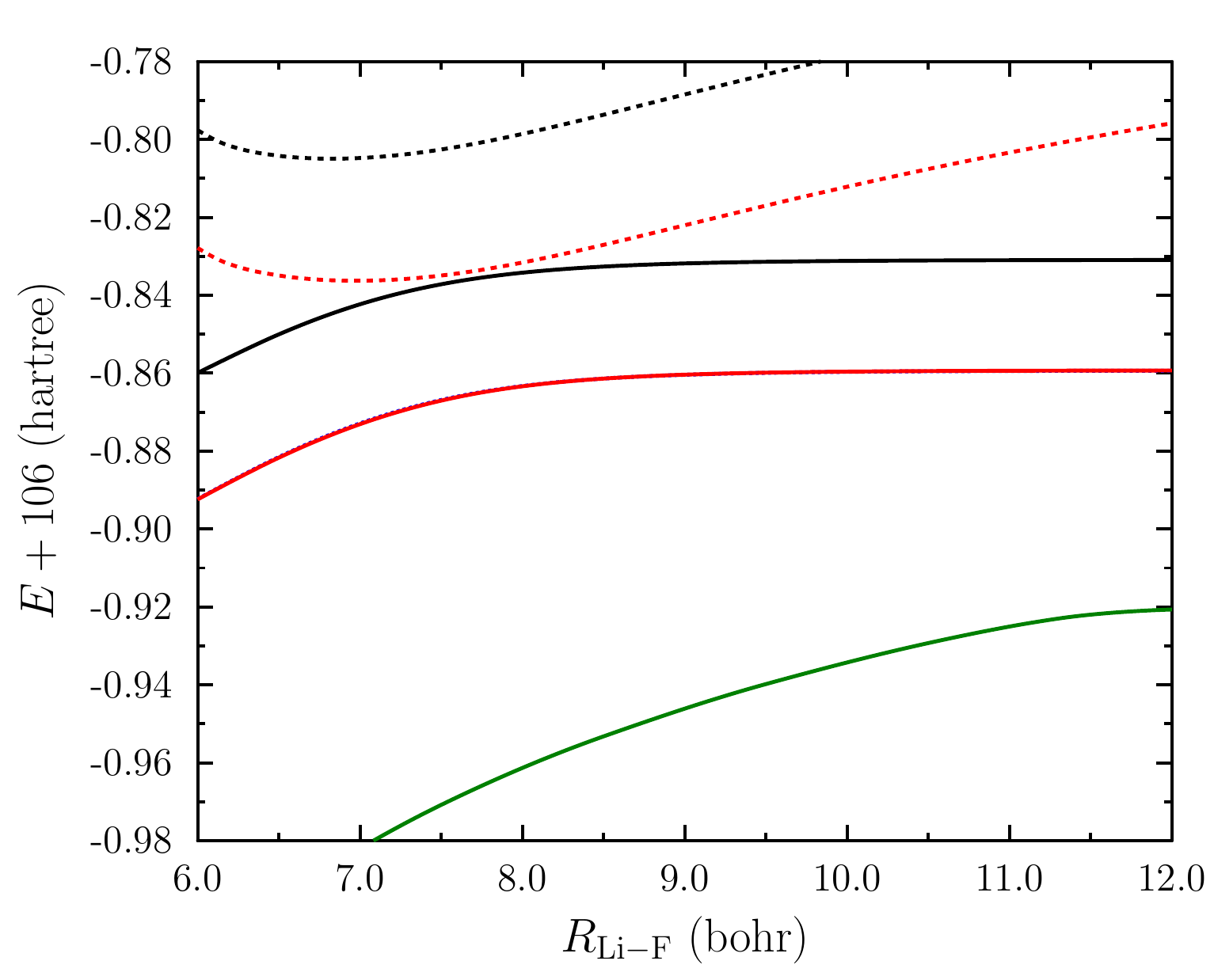}
  \hspace{0.2cm}
  \includegraphics[width=8cm]{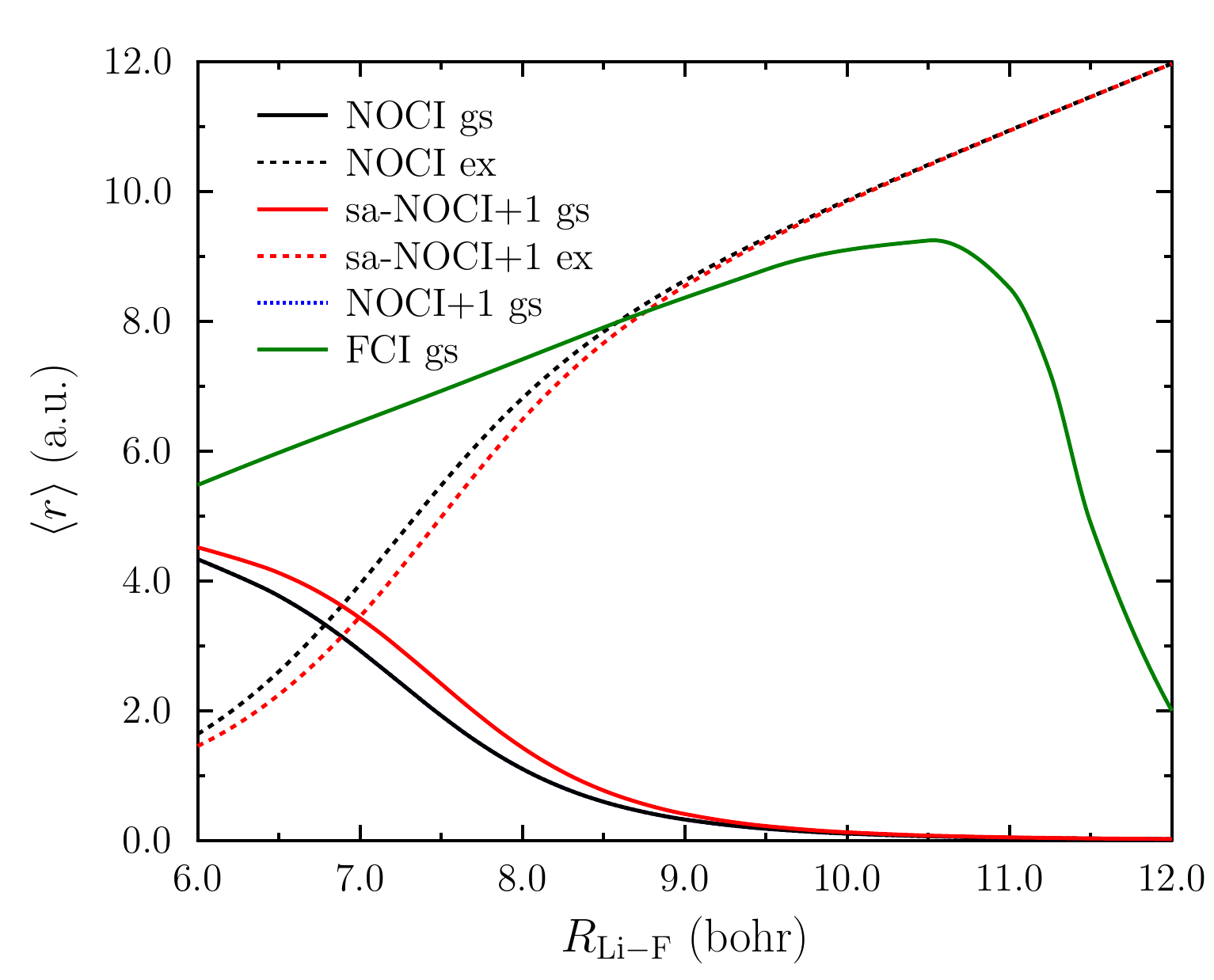}
  \caption{(Left) Ground and excited state energies in LiF with NOCI
    and sa-NOCI+1 (after a single gd step in a state-averaged
    formulation). We also show the g.s. energy obtained with NOCI+1,
    where only the ground state is targeted (the resulting curve is
    nearly indistinguishable from the sa-NOCI+1 g.s. curve). (Right)
    Dipole moment evaluated with the same methods. The NOCI+1
    g.s. curve is in this case nearly indistinguishable from the NOCI
    g.s. curve. Note that after one gd step in sa-NOCI the crossing of
    the dipole moment is shifted by 0.5 bohr or so. The FCI
    g.s. results shown, obtained using a frozen-core approximation,
    are from Ref. \onlinecite{bauschlicher1988}. \label{fig:lif}}
\end{figure*}

We show in the left panel of Fig. \ref{fig:lif} the potential energy
curves of the ground and excited state obtained with NOCI as well as
sa-NOCI+1. Additionally, we show in the right panel the dipole moment
for the ${}^1 \Sigma$ states. As shown in the left panel, only a
fraction of the missing correlation energy is captured by a single gd
step, but significantly better results could be obtained if more steps
were taken (not done in this work). As shown in the right panel, the
FCI ground state dipole moment peaks near $10.5$~bohr or so, at which
point the avoided crossing occurs. NOCI predicts an avoided crossing
near $6.7$~bohr or so (the crossing of the two black curves), a
reflection of the poor quantitative agreement with FCI. After one gd
step, the avoided crossing in sa-NOCI+1 shifts by about $0.5$~bohr or
so in the right direction. While this is only a modest improvement
(consistent with only a fraction of the missing correlation energy
recovered), we still find it encouraging.

It is interesting to compare sa-NOCI+1 with NOCI+1, where a single gd
step is used to improve just the NOCI ground state. In NOCI+1, the
dipole moment of the ground state is nearly identical to that of NOCI
itself. This suggests that the eigenvector obtained after the
diagonalization of the Hamiltonian matrix in sa-NOCI+1 has significant
mixing between the ground and the excited state. It also implies that
the internal contraction used in NOCI+1 ({\em i.e.}, the eigenvector
from NOCI) leads to larger qualitative errors in the ground state
dipole moment versus sa-NOCI+1.

\subsection{Formaldehyde low-energy spectrum}

As a last example we consider the low-energy spectrum of the
formaldehyde molecule. In a recent paper \cite{jimenez2019}, we showed
how a state-averaged resonating Hartree--Fock approach (sa-ResHF) can
yield a good description of the low-energy spectrum of
formaldehyde. Upon revisiting those results, we realized that most of
the states are well described by a single SUHF
determinant\footnote{Naturally, the SUHF determinants describing
  higher energy states do not correspond to the lowest energy SUHF
  solution but rather to higher-energy oners.}, with the only
exception being the $1 \,\, {}^1A_1$ and $3 \,\, {}^1A_1$ states. In
those two states (the ground and $\pi \to \pi^\ast$ states), the
mixing between the ground RHF-like determinant and the excited $\pi
\to \pi^\ast$ SUHF determinant is quite significant, such that the
state-averaged resonating Hartree--Fock description, where the orbital
optimization is done targeting the state-averaged energy, is
required. The calculations in this work therefore use a single SUHF
determinant for each state, except for the $1 \,\, {}^1A_1$ and $3
\,\, {}^1A_1$ states.\footnote{This accounts for the small differences
  between the results here reported and those in
  Ref. \onlinecite{jimenez2019}.} Our calculations use the same
aug-cc-pVDZ basis set used previously. For the triplet states we have
used $m_s = 0$ UHF determinants in SUHF solutions, although we
emphasize that the spin projection was done to a triplet state.

We show in Tab. \ref{tab:formaldehyde} the vertical excitation
energies obtained by sa-ResHF and sa-(sa-ResHF)+1, where in the latter
case we have carried out a single gd step in the state-averaged
formalism described in this work. In this case we have also directly
evaluated the $F_2$ and $F_3$ matrices without an explicit vector
representation of the reference states: the dimension of the Hilbert
space for $m_s=0$ determinants is $2.0 \times 10^{19}$. We note that
the calculations for each symmetry sector were carried out
independently.

As shown in Tab. \ref{tab:formaldehyde}, the ground state energy is
lowered by $>5$~eV after a single gd step. Nonetheless, we only see
small differences (the largest differences are about $0.6$~eV) between
the reference spectrum and that obtained after the gd step. Both of
the ${}^1B_1$ states are shifted by around $0.55$~eV; part of that
shift is likely because the calculations on each symmetry sector were
done independently. If we focus on the relative energy shift within
each symmetry sector, all the shifts are below $0.2$~eV, with the
single outlier being the $2 \,\, {}^1A_1$ state.

It is instructive, for the case of the ${}^1 A_1$ states, to look at
the $S_{1\ast}$ and $H_{1\ast}$ matrices. (Recall that the solution to
the generalized eigenvalue problem using $H_{1\ast}$ and $S_{1\ast}$
yields the low-energy spectrum after one gd step.) They are given by
(with the $H_{1\ast}$ matrix expressed in a.u.)
\begin{align}
  S_{1\ast} &= \, \begin{pmatrix}
    1.0269 &-0.0000 &-0.0020 \\
   -0.0000 & 1.0240 & 0.0001 \\
   -0.0020 & 0.0001 & 1.0275
  \end{pmatrix}, \nonumber \\
  H_{1\ast} &= \, \begin{pmatrix}
    -149.4399 & 0.0046 & 0.3123 \\
    0.0046 & -148.6996 & -0.0116 \\
    0.3123 & -0.0116 & -149.1454
  \end{pmatrix}. \nonumber
\end{align}
The reader may convince himself that the resulting eigenvectors have
considerable mixing between the ground and the $\pi \to \pi^\ast$
states. Note that this is beyond the mixing present in sa-ResHF
itself, as the $S_{1\ast}$ and $H_{1\ast}$ matrices are expressed in
the basis of the orthonomal reference states from sa-ResHF. This
significant mixing implies that the character of those two states is
being adjusted in the presence of the correlation captured by the
single gd step.

\begin{table}[!htb]
  \caption{Vertical excitation energies (in eV) of several low-lying
    singlet and triplet states of formaldehyde evaluated using the
    aug-cc-pVDZ basis set. \label{tab:formaldehyde}}
  \begin{ruledtabular}
    \begin{tabular}{l l r r}
     state & character & sa-ResHF & sa-(sa-ResHF)+1 \\\hline
     singlet states  \\
     $1 \,\, {}^1A_1$ &                           &  0.00\footnote{The ground state energy is $-113.940\,521$~a.u.} &  0.00\footnote{The ground state energy is $-114.127\,074$~a.u.} \\
     $2 \,\, {}^1A_1$ & n $\to$ 3pb$_2$           &  7.74 &  8.35 \\
     $3 \,\, {}^1A_1$ & $\pi$ $\to$ $\pi^\ast$    & 10.47 & 10.32 \\
     $1 \,\, {}^1A_2$ & n $\to$ $\pi^\ast$        &  3.61 &  4.04 \\
     $2 \,\, {}^1A_2$ & n $\to$ 3pb$_1$           &  8.33 &  8.95 \\
     $1 \,\, {}^1B_1$ & $\sigma$ $\to$ $\pi^\ast$ &  9.27 &  9.59 \\
     $1 \,\, {}^1B_2$ & n $\to$ 3sa$_1$           &  6.81 &  7.37 \\
     $2 \,\, {}^1B_2$ & n $\to$ 3pa$_1$           &  7.84 &  8.39 \\[6pt] \hline
     triplet states  \\
     $1 \,\, {}^3A_1$ & $\pi$ $\to$ $\pi^\ast$    &  5.94 &  6.16 \\
     $2 \,\, {}^3A_1$ & n $\to$ 3pb$_2$           &  8.46 &  8.56 \\
     $1 \,\, {}^3A_2$ & n $\to$ $\pi^\ast$        &  4.01 &  3.98 \\
     $2 \,\, {}^3A_2$ & n $\to$ 3pb$_1$           &  9.23 &  9.32 \\
     $1 \,\, {}^3B_1$ & $\sigma$ $\to$ $\pi^\ast$ &  8.98 &  8.96 \\
     $1 \,\, {}^3B_2$ & n $\to$ 3sa$_1$           &  7.52 &  7.57 \\
     $2 \,\, {}^3B_2$ & n $\to$ 3pa$_1$           &  8.52 &  8.54
    \end{tabular}
  \end{ruledtabular}
\end{table}
  
\section{Conclusions}
\label{sec:conclusions}

We have generalized the gradient descent and quasi-Newton algorithms
presented in paper I to the optimization of a low-energy spectrum
instead of just the ground state. The method presented optimizes the
state-averaged energy thereby recovering the exact low-energy spectrum
as the algorithm reaches convergence. The state-averaged energies
along the optimization path are fully expressed in terms of transition
matrix elements among the states in the reference subspace. This
allows us to avoid an explicit vector representation of the
intermediate wavefunctions which is crucial for systems where the
dimension of the Hilbert space becomes intractable. Moreover, the
algorithm defines a systematic approximation to the exact low-energy
spectrum.

We have shown an application of the algorithm using a reference
subspace written as a non-orthogonal configuration interaction in the
case of LiF and formaldehyde. While we only carried out a single step
of the algorithm in those cases, the results can be improved by
carrying out a few more steps or using an improved reference subspace.

Our calculations in LiF and formaldehyde showed that in some cases
there is significant mixing between the states in the reference
subspace in the presence of the missing correlation captured by the
algorithm. That is, even when the reference subspace was deemed as
qualitatively correct, the weights of the reference configurations
adjust as they evolve towards the FCI states. As a consequence of
this, if the method is used to target the ground state exclusively
this can lead to larger qualitative errors compared to the
state-averaged description.

As presented, the method can use other type of reference subspaces
such as complete active space (CAS) or more general
mutli-configurational self-consistent field (MC-SCF) solutions. We
plan to explore the utility of the systematic approximation here
presented using those wavefunctions in the near future.

\begin{acknowledgments}
This work was supported by a generous start-up package from Wesleyan
University.
\end{acknowledgments}

\appendix

\section{BFGS Update}
\label{sec:bfgs}

We discuss in this appendix the form of the search direction $p_1 =
-B_1 g_1$, with $B_1$ constructed from a
Broyden-Fletcher-Goldfarb-Shanno (BFGS)
\cite{broyden,fletcher,goldfarb,shanno} update formula (starting from
$B_0 = I$). Let
\begin{align}
  s_0 &= \, Z_1 - Z_0 = Z_1, \\
  y_0 &= \, g_1 - g_0, 
\end{align}
which yields $s_0= -\sigma_\ast g_0$ and
\begin{equation}
  (y_0)^k_x = H_x^l (\alpha_1-\alpha_0)_l^k + H_x^y H_y^l
  (\beta_1)_l^k.
\end{equation}

Defining $\rho_0 \equiv 1/[(s_0)^x_k (y_0)^k_x]$, the BFGS update
takes the form
\begin{align}
  {}_s^k(B_1)_l^t &= \, {}_s^k(B_0)_l^t \nonumber \\
  &- \, \rho_0 \,\, {}_s^k (B_0)_m^x (y_0)^m_x (s_0)^t_l -
  \rho_0 (s_0)^k_s (y_0)^x_m \,\, {}_x^m (B_0)_l^t \nonumber \\
  &+ \, \rho_0^2 \Big[ \rho^{-1}_0 + (y_0)^x_m \,\, {}_x^m (B_0)_n^y (y_0)^n_y \Big]
  (s_0)_s^k (s_0)^t_l
\end{align}
We now carry an explicit evaluation of $p_1 = -B_1 g_1$. We note that
\begin{align}
  \rho^{-1}_0 &= \, -\sigma_\ast \mathrm{Tr} \left\{ \alpha_0 F_2
    (\alpha_1-\alpha_0) + \alpha_0 F_3 \beta_1 \right\}, \nonumber\\
  [y_0y_0] &\equiv \, (y_0)_k^x (y_0)^k_x \nonumber \\
  &=\, \mathrm{Tr} \left\{ (\alpha_1-\alpha_0) F_2 (\alpha_1-\alpha_0) +
  (\alpha_1-\alpha_0) F_3 \beta_1 \right. \nonumber \\
  &+\, \left. \beta_1 F_3 (\alpha_1-\alpha_0) + \beta_1 F_4 \beta_1
  \right\}, \nonumber \\
  [s_0g_1] &\equiv \, (s_0)_k^x (g_1)^k_x \nonumber \\
  &= \, -\sigma_\ast \mathrm{Tr} \left\{ \alpha_0 F_2 \alpha_1 +
  \alpha_0 F_3 \beta_1 \right\}, \nonumber \\
  [y_0g_1] &\equiv \, (y_0)_k^x (g_1)^k_x \nonumber \\
  &= \,  \mathrm{Tr} \left\{ (\alpha_1-\alpha_0) F_2 \alpha_1
  + (\alpha_1-\alpha_0) F_3 \beta_1 \right. \nonumber \\
  &+ \, \left. \beta_1 F_3 \alpha_1 + \beta_1 F_4 \beta_1
  \right\}. \nonumber
\end{align}
Therefore, $p_1$ takes the form
\begin{equation}
  (p_1)^k_x = H_x^l (\alpha'_1)^k_l + H_x^y H_y^l (\beta'_1)^k_l ,
\end{equation}
with
\begin{align}
  \alpha'_1 &=\, \alpha_1 - \sigma_\ast \rho_0^2 (\rho_0^{-1} +
         [y_0y_0]) [s_0g_1] \alpha_0 \nonumber \\
  &- \, \rho_0 [s_0g_1] (\alpha_1-\alpha_0) + \sigma_\ast \rho_0
         [y_0g_1] \alpha_0, \\
  \beta'_1 &=\, \beta_1 - \rho_0 [s_0g_1] \beta_1.
\end{align}

\bibliography{paper2}

\begin{thebibliography}{14}
\expandafter\ifx\csname natexlab\endcsname\relax\def\natexlab#1{#1}\fi
\expandafter\ifx\csname bibnamefont\endcsname\relax
  \def\bibnamefont#1{#1}\fi
\expandafter\ifx\csname bibfnamefont\endcsname\relax
  \def\bibfnamefont#1{#1}\fi
\expandafter\ifx\csname citenamefont\endcsname\relax
  \def\citenamefont#1{#1}\fi
\expandafter\ifx\csname url\endcsname\relax
  \def\url#1{\texttt{#1}}\fi
\expandafter\ifx\csname urlprefix\endcsname\relax\def\urlprefix{URL }\fi
\providecommand{\bibinfo}[2]{#2}
\providecommand{\eprint}[2][]{\url{#2}}

\bibitem[{\citenamefont{Szabo and Ostlund}(1989)}]{szabo_ostlund}
\bibinfo{author}{\bibfnamefont{A.}~\bibnamefont{Szabo}} \bibnamefont{and}
  \bibinfo{author}{\bibfnamefont{N.~S.} \bibnamefont{Ostlund}},
  \emph{\bibinfo{title}{Modern Quantum Chemistry: Introduction to Advanced
  Electronic Structure Theory}} (\bibinfo{publisher}{McGraw-Hill},
  \bibinfo{address}{New York}, \bibinfo{year}{1989}).

\bibitem[{\citenamefont{{Jim\'enez}-Hoyos}(2020)}]{jimenez2020}
\bibinfo{author}{\bibfnamefont{C.~A.} \bibnamefont{{Jim\'enez}-Hoyos}},
  \emph{\bibinfo{title}{Approaching the full configuration interaction ground
  state from an arbitrary wavefunction with gradient descent and quasi-{N}ewton
  algorithms}} (\bibinfo{year}{2020}), \eprint{arXiv:2010.02027}.

\bibitem[{\citenamefont{Sundstrom and {Head-Gordon}}(2014)}]{sundstrom2014}
\bibinfo{author}{\bibfnamefont{E.~J.} \bibnamefont{Sundstrom}}
  \bibnamefont{and}
  \bibinfo{author}{\bibfnamefont{M.}~\bibnamefont{{Head-Gordon}}},
  \bibinfo{journal}{J. Chem. Phys.} \textbf{\bibinfo{volume}{140}},
  \bibinfo{pages}{114103} (\bibinfo{year}{2014}).

\bibitem[{\citenamefont{Jensen et~al.}(2018)\citenamefont{Jensen, Benson,
  Cardamone, and Thom}}]{thom2018}
\bibinfo{author}{\bibfnamefont{K.~T.} \bibnamefont{Jensen}},
  \bibinfo{author}{\bibfnamefont{R.~L.} \bibnamefont{Benson}},
  \bibinfo{author}{\bibfnamefont{S.}~\bibnamefont{Cardamone}},
  \bibnamefont{and} \bibinfo{author}{\bibfnamefont{A.~J.~W.}
  \bibnamefont{Thom}}, \bibinfo{journal}{J. Chem. Theory Comput.}
  \textbf{\bibinfo{volume}{14}}, \bibinfo{pages}{4629} (\bibinfo{year}{2018}).

\bibitem[{\citenamefont{{Jim\'enez}-Hoyos
  et~al.}(2012)\citenamefont{{Jim\'enez}-Hoyos, Henderson, Tsuchimochi, and
  Scuseria}}]{jimenez2012}
\bibinfo{author}{\bibfnamefont{C.~A.} \bibnamefont{{Jim\'enez}-Hoyos}},
  \bibinfo{author}{\bibfnamefont{T.~M.} \bibnamefont{Henderson}},
  \bibinfo{author}{\bibfnamefont{T.}~\bibnamefont{Tsuchimochi}},
  \bibnamefont{and} \bibinfo{author}{\bibfnamefont{G.~E.}
  \bibnamefont{Scuseria}}, \bibinfo{journal}{J. Chem. Phys.}
  \textbf{\bibinfo{volume}{136}}, \bibinfo{pages}{164109}
  (\bibinfo{year}{2012}).

\bibitem[{\citenamefont{Nite and {Jim\'enez}-{H}oyos}(2019)}]{jimenez2019}
\bibinfo{author}{\bibfnamefont{J.}~\bibnamefont{Nite}} \bibnamefont{and}
  \bibinfo{author}{\bibfnamefont{C.~A.} \bibnamefont{{Jim\'enez}-{H}oyos}},
  \bibinfo{journal}{J. Chem. Theory Comput.} \textbf{\bibinfo{volume}{15}},
  \bibinfo{pages}{5343} (\bibinfo{year}{2019}).

\bibitem[{\citenamefont{Nocedal and Wright}(2006)}]{nocedal}
\bibinfo{author}{\bibfnamefont{J.}~\bibnamefont{Nocedal}} \bibnamefont{and}
  \bibinfo{author}{\bibfnamefont{S.~J.} \bibnamefont{Wright}},
  \emph{\bibinfo{title}{Numerical Optimization}}
  (\bibinfo{publisher}{Springer}, \bibinfo{address}{New York},
  \bibinfo{year}{2006}), \bibinfo{edition}{2nd} ed.

\bibitem[{\citenamefont{Broyden}(1970)}]{broyden}
\bibinfo{author}{\bibfnamefont{C.~G.} \bibnamefont{Broyden}},
  \bibinfo{journal}{{IMA} J. Appl. Math.} \textbf{\bibinfo{volume}{6}},
  \bibinfo{pages}{76} (\bibinfo{year}{1970}).

\bibitem[{\citenamefont{Fletcher}(1970)}]{fletcher}
\bibinfo{author}{\bibfnamefont{R.}~\bibnamefont{Fletcher}},
  \bibinfo{journal}{Comput. J.} \textbf{\bibinfo{volume}{13}},
  \bibinfo{pages}{317} (\bibinfo{year}{1970}).

\bibitem[{\citenamefont{Goldfarb}(1970)}]{goldfarb}
\bibinfo{author}{\bibfnamefont{D.}~\bibnamefont{Goldfarb}},
  \bibinfo{journal}{Math. Comp.} \textbf{\bibinfo{volume}{24}},
  \bibinfo{pages}{23} (\bibinfo{year}{1970}).

\bibitem[{\citenamefont{Shanno}(1970)}]{shanno}
\bibinfo{author}{\bibfnamefont{D.~F.} \bibnamefont{Shanno}},
  \bibinfo{journal}{Math. Comp.} \textbf{\bibinfo{volume}{24}},
  \bibinfo{pages}{647} (\bibinfo{year}{1970}).

\bibitem[{\citenamefont{Qiu et~al.}(2017)\citenamefont{Qiu, Henderson, and
  Scuseria}}]{scuseria2017}
\bibinfo{author}{\bibfnamefont{Y.}~\bibnamefont{Qiu}},
  \bibinfo{author}{\bibfnamefont{T.~M.} \bibnamefont{Henderson}},
  \bibnamefont{and} \bibinfo{author}{\bibfnamefont{G.~E.}
  \bibnamefont{Scuseria}}, \bibinfo{journal}{J. Chem. Phys.}
  \textbf{\bibinfo{volume}{146}}, \bibinfo{pages}{184105}
  (\bibinfo{year}{2017}).

\bibitem[{\citenamefont{Bauschlicher and Langhoff}(1988)}]{bauschlicher1988}
\bibinfo{author}{\bibfnamefont{C.~W.} \bibnamefont{Bauschlicher}}
  \bibnamefont{and} \bibinfo{author}{\bibfnamefont{S.~R.}
  \bibnamefont{Langhoff}}, \bibinfo{journal}{J. Chem. Phys.}
  \textbf{\bibinfo{volume}{89}}, \bibinfo{pages}{4246} (\bibinfo{year}{1988}).

\bibitem[{\citenamefont{Nite and {Jim\'enez}-Hoyos}(2019)}]{jimenez2019b}
\bibinfo{author}{\bibfnamefont{J.}~\bibnamefont{Nite}} \bibnamefont{and}
  \bibinfo{author}{\bibfnamefont{C.~A.} \bibnamefont{{Jim\'enez}-Hoyos}},
  \emph{\bibinfo{title}{Efficient multi-configurational wavefunction method
  with dynamical correlation using non-orthogonal configuration interaction
  singles and doubles ({NOCISD})}} (\bibinfo{year}{2019}),
  \eprint{ChemRxiv:11369641}.

\end{thebibliography}

\end{document}